\documentclass[11pt]{article}
\usepackage{moriond,epsfig,here}

\def\be{\begin{equation}}
\def\ee{\end{equation}}
\def\bea{\begin{eqnarray}}
\def\eea{\end{eqnarray}}

\begin{document}
\vspace*{4cm}
\title{CKM fits as of winter 2009 and sensitivity to New Physics}

\author{Vincent Tisserand, on behalf of the CKMfitter group }

\address{LAPP (UMR5814), Universit\'e de Savoie, CNRS/IN2P3, Annecy-le-Vieux, France}

\maketitle\abstracts{We present the status of the CKM matrix
parameters in the framework of the Standard Model. We perform a
model independent analysis to set constraints on additional
effective parameters accounting for possible New Physics effects
and to evaluate the present allowed space for these effects both
from $B_d$ and $B_s$ mesons.}

The unitary Cabibbo-Kobayashi-Maskawa (CKM) matrix~\cite{ref:CKM}
describes the mixing of the quark flavors within the framework of
the Standard Model (SM). Profs. Kobayashi and Maskawa have just
been awarded the Nobel prize for their early 70's work on such a
$3\times3$ (3 quark generations) unitary matrix that accounts for
violation of CP symmetry through Electro-Weak (EW) couplings. It
has 4 real parameters, among which one single non-vanishing phase.
We employ an exact Wolfenstein-like
parametrization~\cite{ref:Wolfenstein,ref:thePapII} that describes
the strong hierarchy in these couplings where unitarity holds to
an arbitrary power of the Cabibbo angle $\lambda=\sin(\theta_C)$,
it is also re-phasing invariant:
\begin{eqnarray}
{\lambda^2=\frac{|V_{us}|^2}{|V_{ud}|^2+|V_{us}|^2}}, \qquad
{A^2\lambda^4 =\frac{|V_{cb}|^2}{|V_{ud}|^2+|V_{us}|^2}}, \qquad
\bar\rho+i\bar\eta=-\frac{V_{ud}V_{ub}^*}{V_{cd}V_{cb}^*}.
\nonumber
\end{eqnarray}
The parameter $\lambda$ is accurately determined (at 0.3~\% level)
from super-allowed nuclear transitions ($|V_{ud}|$) and in
semi-leptonic kaon decays ($|V_{us}|$). The parameter $A$
($|V_{cb}|$) is measured from charmed $B$ semi-leptonic decays
with an accuracy at the level of 3~\%. The apex of the Unitary
Triangle (UT), i.e. the complex number ($\bar\rho+i\bar\eta$), is
less constrained.

The accurate measurement of these parameters and especially of the
UT sides and angles, possibly in a redundant way, allows to check
the consistency of the Kobayashi-Maskawa (KM) mechanism within the
SM. Any significant departure could suggest contributions from New
Physics (NP). The challenge, both for experimentalists and
theorists, is that precise extraction of observables related to
these EW parameters is complicated by the presence of strong
interaction effects.

We perform a global fit to the CKM parameters within a frequentist
approach including a specific treatment to deal with theoretical
uncertainties (i.e. flat likelihood {\it \`a la
Rfit})\cite{ref:thePapII}, where we only use the observables from
$K$ and $B$ sectors on which we have a good theoretical control,
to avoid to claim pseudo departures from SM.
Table~\ref{tab:inputs} displays the various key ingredients used
(more details on the world averages (WA) exp. and theo. inputs and
related references are given at~\cite{ref:thePapII}). Among all
these observables, only the branching ratio (BR) of the
$B^+\to\rho^+\rho^0$ channel updated by the BaBar
collaboration~\cite{ref:BaBarrhorho} is a new input since our last
summer 2008 update.

\begin{table}[h]
{\scriptsize
\begin{center}
\begin{tabular}{|c|c|c|}
\hline
Phys. params.   & Experim. input  & Theory method/ingredient \\
\hline \hline
$|V_{ud}|$    &  super-allowed $\beta$ decays  & Towner and Hardy (08)   \\
$|V_{us}|$    &  $K_{l3}$ SL kaon decays  (WA, Flavianet~\cite{ref:Flavianet})   & $f_+^{K\pi}(0)=0.964(5)$ (RBC-UKQCD (07))   \\
$|V_{cb}|$ &   $B\to X_c l \nu$ (HFAG~\cite{ref:HFAG}: excl. + incl.) & $40.59(38)(58)\times 10^{-3}$ (FF and/or OPE)  \\
$|V_{ub}|_{SL}$   &  $B\to X_u l \nu$ (HFAG~\cite{ref:HFAG}: excl. + incl.)   &$3.87(9)(46)\times 10^{-3}$  (FF and/or OPE)  \\
& & and own syst. treatment \\
$|V_{ub}|_{lept.}$  & BR($B^+\to\tau^+\nu$) annihilation
(B-factories) &
[$f_{B_s}=228(3)(17)$ MeV, $f_{B_s}/f_{B_d}=1.196(8)(23)]$ \\
$\Delta m_s$  & $B_s - \bar B_s$ mixing (CDF II)& $\Delta B$=2
amp.  [$\hat B_s=1.23(3)(5)$, $f_{B_s}$, $\bar m_t$ , $\eta_{B}$] \\
$\Delta m_d$ &  $B_d - \bar B_d$ mixing (HFAG~\cite{ref:HFAG}) &
$\Delta B$=2
amp.  [$\hat B_{B_s}/\hat B_{B_d}=1.05(2)(5)$, \\
 & & $f_{B_s}/f_{B_d}$, $\eta_{B}$]  \\
\hline \hline $|\varepsilon_K|$ & $K\bar K$ mixing (PDG
08~\cite{ref:PDG08}: KLOE, NA48, KTeV...) & $\Delta S$=2 amp.
[$B_K=0.721(5)(40)$, $\eta_{cc}$ , $\eta_{ct}$, $\eta_{tt}$] \\
$\beta/\phi_1$ & Charmonium $B$ decays  (HFAG~\cite{ref:HFAG}) & -    \\
$\alpha/\phi_2$ & $B\to\pi\pi,\rho\rho,\rho\pi$ (B-factories: rates + asym.)   & Isopsin $SU(2)$ (Gronau, London (90))    \\
$\gamma/\phi_3$ &  $B^-\to D^{(*)}K^{(*)-}$ (B-factories: rates + asym.) & GLW/ADS/GGSZ   \\
 \hline
\end{tabular}
\caption{ Various relevant inputs to the CKMfitter global fit.
Many LQCD inputs in these table are from our own average (see
text). The upper (lower) part of the table corresponds to CP
conserving (violating) parameters. \label{tab:inputs}}
\end{center}
} 
\end{table}

Several hadronic inputs are mandatory for the fits. They mainly
limit the precision on the determination of the observables
involving processes with loops such as  $\Delta m_d$, $\Delta
m_s$, $|\varepsilon_K|$, and also the tree decay
$B^+\to\tau^+\nu$. The hadronic contributions to $K_{l3}$ decay
are surprisingly under excellent control. We mostly rely on
lattice QCD (LQCD) simulations to estimate these quantities, since
the accuracy of such first-principle computations can be improved
in a controlled way (at least in principle). The presence of
results from different collaborations with various statistics and
systematics makes it all the more necessary to combine them in a
careful and reproducible way. It has been pointed
out\cite{ref:Ligeti} that {\it ``if experts cannot agree, it is
unlikely the rest of the community would believe a claim of NP"}.
Therefore we have recently set up our own average of these
results~\footnote{ We apply the averaging
procedure~\cite{ref:thePapII}:
\begin{itemize}
\item First of all, we collect the relevant calculations of the
quantity that we are interested in: we take only unquenched
results with 2 or 2+1 dynamical fermions, from published papers or
proceedings. In these results, we separate the error estimates
into a Gaussian part and a flat  part (Rfit). The Gaussian part
should collect the uncertainties from purely statistical origin,
but also the systematics that can be controlled and treated in a
similar way (e.g., interpolation or fitting in some cases). The
remaining systematics constitute the Rfit error. If there are
several sources of error in the Rfit category, we add them
linearly, keeping in mind that in many papers in the literature,
this combination is done in quadrature and the splitting between
different sources is not published. If Rfit is taken stricto sensu
and the  individual likelihoods are combined in the usual way (by
multiplication), the final uncertainty can be underestimated, in
particular in the case of marginally compatible values.

\item We correct this effect by adopting the following averaging
recipe. We first combine the Gaussian uncertainties by combining
likelihoods restricted to their Gaussian part. Then we assign to
this combination the smallest of the individual Rfit
uncertainties. The underlying idea is twofold: (1) the present
state of art cannot allow us to reach a better theoretical
accuracy than the best of all estimates, and (2) this best
estimate should not be penalized by less precise methods (as it
would happen be the case if one would take the dispersion of the
individual central values as a guess of the combined theoretical
uncertainty). It should be stressed that the concept of a
theoretical uncertainty is ill-defined, and the combination of
them even more. Thus our approach is only one among the
alternatives that can be found in the literature. In contrast to
some of the latter, ours is algorithmic and can be reproduced. We
found a very good agreement between our previous inputs (taken
from lattice reviews) and our current set (obtained from the above
recipe).
\end{itemize}}.

\begin{figure}[h]
\begin{center}
{\hspace{-1cm}
\epsfig{file=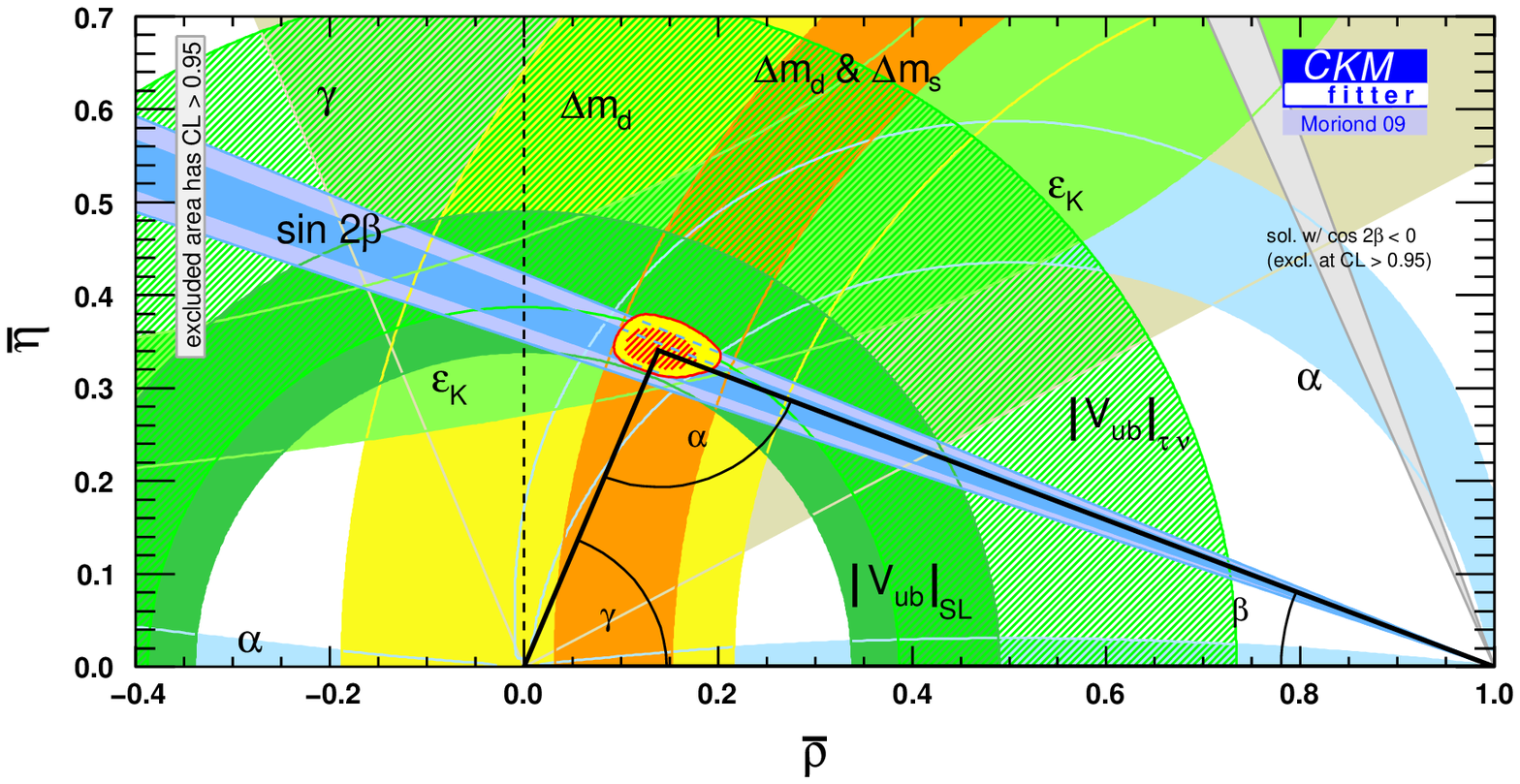,height=6.15cm,width=0.6\linewidth}}
{\hspace{-0.5cm}
\epsfig{file=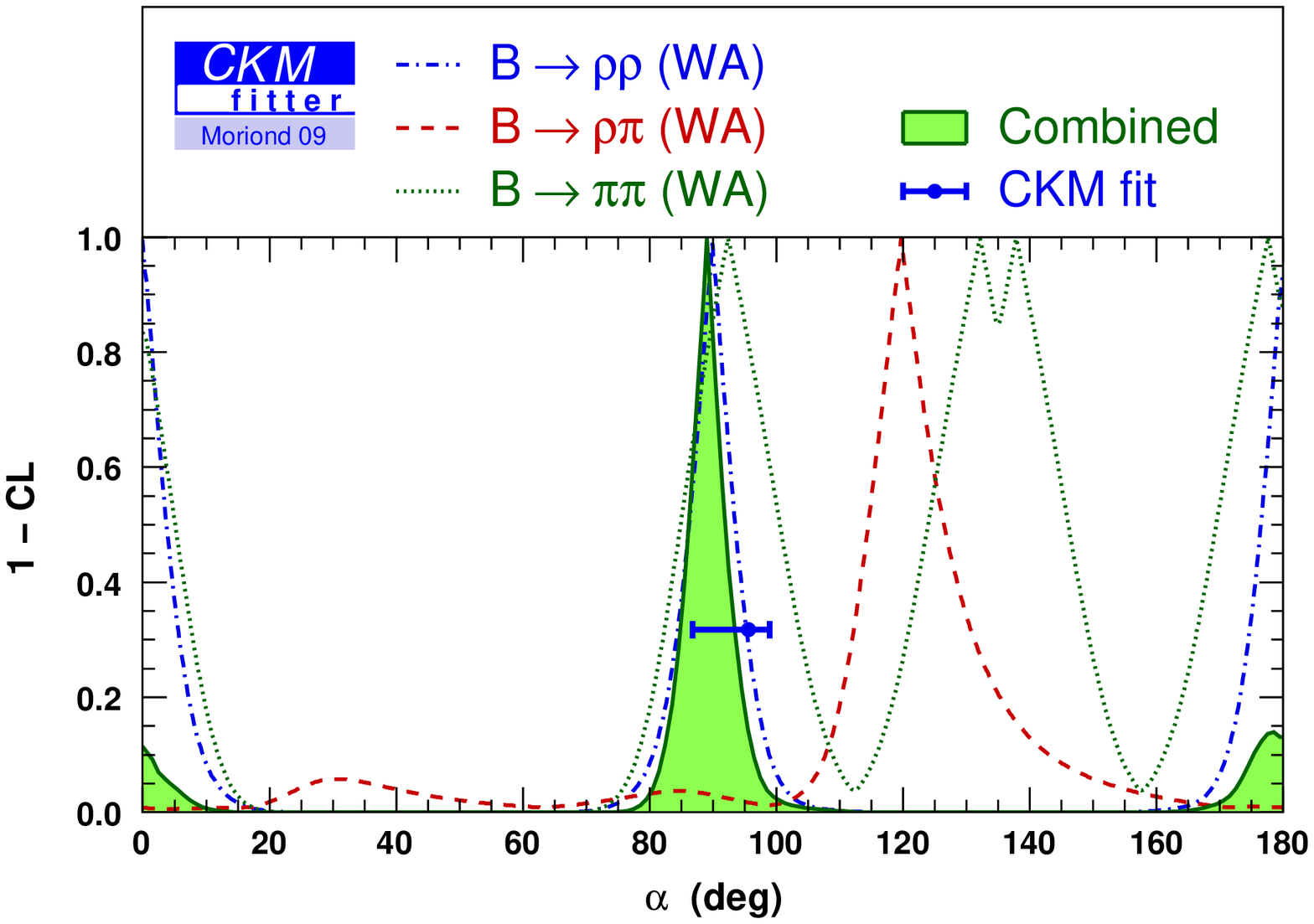,height=6cm,width=0.45\linewidth}}
 \vspace{-0.25cm}
\caption{95~\% CL individual and global constraints in the
($\bar\rho$,$\bar\eta$) plane from the global CKM fit (Left). The
red hashed region of the global combination corresponds to 68~\%
CL. CL profile for $\alpha$ with the present world average of the
3 $B\to\pi\pi,\rho\rho,\rho\pi$ channels (Right).
\label{fig:global}}
\end{center}
\end{figure}

Figure~\ref{fig:global} (Left) shows the global CKM fit results in
the ($\bar\rho$,$\bar\eta$) plane. The CKM parameters are:
$A=0.8116^{+0.0097}_{-0.0241}$, $\lambda=0.22521\pm0.00082$,
$\bar\rho=0.139^{+0.025}_{-0.027}$, and
$\bar\eta=0.341^{+0.016}_{-0.015}$. A good overall consistency at
95~\% CL is seen, probing the fact that the KM mechanism is at
work for CP violation and dominant in $B$ decays. It is also
visible that there is a tension between the measurement of
$\sin(2\beta)$ from charmonium $B$ decays and the determination of
$|V_{ub}|$ from the decay $B^+\to\tau^+\nu$. When removing one of
the last paramaters from the global fit, the $\chi^2$ at minimum
drops respectively by 2.3 and 2.4~$\sigma$.

This tension is mainly originated from the recent
BR($B^+\to\tau^+\nu$) measurements by BaBar and
Belle~\cite{ref:thePapII,ref:ODeschamps}. All these measurements
are consistent and their WA is $(1.73\pm0.35)\times10^{-4}$, while
our global CKM fit predicts it to be at a lower value of
$(0.80^{+0.15}_{-0.09})\times10^{-4}$. Such a higher BR is not
necessarily accommodated for by models with 2 Higgs boson
doublets~\cite{ref:Monteil} (2HDM). In addition one can see on
Fig.~\ref{fig:global} that both semi-leptonic and purely leptonic
$B$ decays $|V_{ub}|$ determinations agree pretty well. In between
the 2 semi-leptonic methods $\sin(2 \beta)$ prefers the exclusive
one, while the inclusive one is still compatible in the CKMfitter
approach. Doing the computation of the ratio of the  BR of this
$B$ annihilation decay over the mixing parameter $\Delta m_d$
removes the dependance to the decay factor $f_{B_d}$. The
combination of these 2 constraints releases therefore partially
some LQCD related uncertainties and gives a direct access to the
parameter $B_{B_d}$~\cite{ref:ODeschamps}. When doing so we obtain
the value $B_{B_d}=1.18\pm0.14$ that is 2.7~$\sigma$ away from the
CKM global fit: $0.52^{+0.15}_{-0.11}$. The tension arising from
the BR($B^+\to\tau^+\nu$) is clearly not yet an evidence for NP,
but it motivates more accurate measurements at BaBar and Belle and
at possible future super-B factories.

It has been suggested~\cite{ref:BurasGuad} that the recent LQCD
improvements in the determination of the parameter $\hat B_K$
alights a so far neglected additional multiplicative factor in the
determination of the parameter $|\varepsilon_K|$, this is the so
called $\kappa_\varepsilon$ parameter computed and estimated to be
equal to $0.92\pm0.02$ . This factor accounts for CP violation
effects in $K-\bar K$ mixing and may hint for CP violation
contributions originated from NP. The computed value of
$|\varepsilon_K|$ from this recent work and within the SM is
$(1.78\pm0.25)\times10^{-3}$, while the current experimental
WA~\cite{ref:PDG08} is $(2.229\pm0.10)\times10^{-3}$. This
suggests an additional tension at the level of 2~$\sigma$ mainly
with respect to $\sin(2\beta)$. Our fit\cite{ref:Monteil}, even
while accounting for $\kappa_\varepsilon$, shows that the
uncertainty of $|\varepsilon_K|$ is rather likely to be of the
order of $0.5\times10^{-3}$. This tension arises while dealing
with pure convoluted Gaussian uncertainties for all the parameters
and including  all the uncertainties on LQCD computations, that
are obviously not overwhelmed by statistical effects. It therefore
vanishes while using the Rfit procedure.

Figure~\ref{fig:global} (Right) shows that the angle $\alpha$ is
now determined with a good accuracy, at the level of 5~\% or less:
$\alpha=(89.0^{+4.4}_{-4.2})^\circ$, while the angle $\beta$ is
measured within a precision of 4~\%. The isospin analysis on the
$\rho\rho$ channels almost fully drives it. It is in excellent
agreement with the global fit $(95.6^{+3.3}_{-8.8})^\circ$
(without the related measurement in the fit) and the uncertainties
have been reduced by more than 20~\% with respect to last summer.
This is due to the new measurement on the BR($B^+\to\rho^+\rho^0$)
by BaBar~\cite{ref:BaBarrhorho} that dominates the WA for this
observable. It has increased from $(18.2\pm3.0)\times10^{-6}$ up
to $(24.0\pm1.9)\times10^{-6}$. In the $\rho\rho$ system, the
Penguin to Tree amplitude ratio is much more favorable than in the
case of charmless $B$ decays to $\rho\pi$ and
$\pi\pi$~\cite{ref:thePapII,ref:ZupanetAl}, allowing therefore a
relatively smaller $|\Delta\alpha|$ isospin bound.

The BR of both channels $\rho^+\rho^0$ and $\rho^+\rho^-$ are now
very similar~\cite{ref:HFAG} and almost 25 times as big as that of
$\rho^0\rho^0$ (the Penguin transition), the $B$ and $\bar B$
related isospin amplitudes triangles are basically flat and do not
close, i.e. for $B$ : $|A^{+-}|/\sqrt{2} + |A^{00}|<|A^{+0}|$ (but
this is still consistent within uncertainties). As a consequence
the mirror solutions that possibly arise while experimentally
measuring the effective angle $\alpha_{eff}$ (Penguin dilution),
are degenerated into a single peak. As it can be seen on
Fig.~\ref{fig:global} the expected 8-fold ambiguities from the
isospin analysis degenerate into the only 4 $\Delta\alpha$
geometric solutions, in the vicinity of $0^\circ$, $\pm~90^\circ$,
and $\pm~180^\circ$.

The isospin analysis for the $\rho\rho$ system is performed
using~\cite{ref:HFAG} the 3 BRs, time-dependent CP-asymmetry
parameters $C^{+-}$, $S^{+-}$, $C^{00}$, and $C^{00}$, and the 3
longitudinal fractions ($f_L$) of these $VV$ channels that are not
stricto-sensu CP-eigenstates, thought the $f_L$ are very close to
1 which eases the analysis. The $\alpha$ angle is determined to be
$(89.9\pm5.4)^\circ$ and the isospin bound $\Delta\alpha$ close to
$0^\circ$ with a good accuracy: $(1.4\pm3.7)^\circ$ (at summer
time we had: $\alpha=(90.9^{+6.7}_{-14.9})^\circ$). To test what
is the expected uncertainties for this measurement, we have
performed 1000 pseudo experiments (toys). We have generated the
above experimental observables with $\pm 1~\sigma$  around their
best fitted value (from global fit), where the $\sigma$ are the
currently measured uncertainties. We measure that the average
expected uncertainty is $7.5^\circ$, slightly higher than the
$5.4^\circ$ that we measure. The uncertainty distribution has a
long tail up to about $20^\circ$, it corresponds to revival of
pseudo mirror solutions, above the $1~\sigma$ CL($\alpha$). About
34~\% of the toys where isospin triangles close and have similar
uncertainties or higher than that of last summer configuration.
This is a message for future experiments, such as LHCb, that
better uncertainties of the various $\rho\rho$ observables may not
necessarily  lead to better accuracy on $\alpha$.

Due to the reached precision, it is legitimate to investigate for
possible isospin breaking effects~\cite{ref:ZupanetAl} beyond the
Gronau-London $SU(2)$ method. Not all the breaking effects can be
calculated at present, but we can list a few of them: the $u$ and
$d$ quarks have different electric charges and masses (breaking of
the order: $(m_u-m_d)/\Lambda_{QCD}\sim 1~\%$), the isospin
transitions $\Delta I=5/2$ may be no more negligible, we may need
to extend the basis of EW-Penguin operators: $Q_{7,...,10}$
($\Delta\alpha_{EWP}\sim 1.5^\circ$), the mass and isospin
eigenstates are different ($\rho-\omega$ mixing at the level of
$2~\%$), the $\rho$ natural width is large enough such that $I=1$
contributions are possible (${\cal O}{
(\Gamma^2_\rho/m^2_\rho)\sim 4~\%)}$ ... There are possible ways
out such as exploiting the $B^+ \to K^\star\rho^+$ channels
through $SU(3)$ constraints. In order to break the triangle
closure we apply the procedure as described
in~\cite{ref:thePapII}. The amplitudes  $A^{+0}$ and $\bar A^{+0}$
are corrected by additional Tree ($\Delta_T$) and Penguin
($\Delta_P$) contributions weighted as: $\sqrt{2}\Delta
A^{+0}=V_{ud}V^\star_{ub} \Delta_T T^{+-} + V_{td}V^\star_{tb}
\Delta_P P^{+-}$ (the strong phases are set arbitrarily). We
tested $|\Delta A^{+0}|$ as big as 4, 10, and $15~\%$. The 2 first
corrections break $SU(2)$ at $90^\circ$ and restore it in the
vicinity of $0^\circ$, while the largest is needed to restore it
at the $\alpha$ SM solution. Anyway when combining the $\pi\pi$
and $\rho\pi$ the determination on $\alpha$ is mostly unaffected
at $1~\sigma$ CL.

We have updated~\cite{ref:thePapII} the constraint on
$|V_{td}/V_{ts}|$ accessible through the ratio of branching ratio
for $B\to V\gamma$ decays, where $V$ holds respectively for
$(\rho,\omega)$ and $K^\star$ vector mesons. These penguins
processes complement the box diagrams involved in the measurement
of $\Delta m_{(d,s)}$. Any inconsistency in between the 2
approaches would teach us in which direction to look for NP. We
use the parametrization for hadronic effects as described
in~\cite{ref:Descotes}. The sophisticated description of the
amplitudes has non trivial sensitivity to the CKM parameters. Our
new analysis benefits from the recent updated BR measurements of
all of the above decays~\cite{ref:HFAG}. The improvement is such
that at 95~\% CL these new measurements constrain the
($\bar\rho$,$\bar\eta$) plane as accurately as $\Delta m_d$ alone,
and at 68~\% CL they have similar precision as that from $\Delta
m_{(d,s)}$ at 95~\% CL.

There has been a standing issue due to apparently non SM BR
measurements for leptonic decays of $D_s$
mesons~\cite{ref:Descotes,ref:Kronfeld}, by the B-factory and the
CLEO-c experiments. These decays give access to the measurement of
the decay parameter $f_{Ds}$ and to $|V_{cs}|$. The charm sector,
where $m_c \sim \Lambda_{QCD}$, is an ideal laboratory to validate
LQCD against experiment. The recent most accurate BR measurements
by CLEO-c~\cite{ref:Rubin} on annihilation decays $D_s \to
(\tau,\mu)\nu$ allow to compute $f_{Ds}=(259.5 \pm 6.6 \pm
3.1)$~MeV, while our average on LQCD results is $(246.3 \pm 1.2
\pm 5.3)$~MeV. There is still some discrepancy at the 2~$\sigma$
level, but it is almost twice as less as what it used to be.
Converting this into a  $|V_{cs}|$ determination and averaging
CLEO-c and LQCD measurements of $f_{Ds}$, one computes
$|V_{cs}|=1.027 \pm 0.051$, in good agreement with the global fit
that yields $0.97347\pm0.00019$. This comparison alighted a
2~$\sigma$ tension  one year ago and the measurements led to a
unitarity violation of the CKM matrix~\cite{ref:Descotes}.

We also updated the constraint from the measured BR of the $K^+
\to \pi^+ \nu \bar \nu$ rare decay, for which a recent update of
the E789 and E949 experiments has been done with 5 signal
candidates~\cite{ref:Kpinunubar}. We parameterize the BR using the
calculations by Brod and Gorbahn at NLO QED-QCD and accounting for
EW corrections to the charm quark contribution. The global fit
predicts BR=$({0.811^{+0.027}_{-0.021}}_{exp.} \pm
0.096_{theo.})\times10^{-10}$ while the experiments measure
$(1.73^{+1.15}_{-1.05})\times10^{-10}$. The agreement is good and
the  constraint in the ($\bar\rho$,$\bar\eta$) plane is such that
in the vicinity of the point (1,0) a non negligible area is
forbidden at 95~\% CL for the first time. This effect clearly
motivates a ${\cal O}(100)$ signal event experiment, such as the
future NA62.

Finally we reiterate~\cite{ref:NiersteLenz,ref:ODeschamps} the
analysis to compute the constraints set on NP from
$B_{q=d,s}$-meson mixing. We consider that NP only affects the
short distance part of the $\Delta B=2$ transitions. In addition
we assume that the tree-level mediated decays proceeding through a
Four Flavor Change get only SM contributions (SM4FC hypothesis:
$b\to q_i {\bar q}_j q_k$ ($i \neq j \neq k$)), the observables
$|V_{ij}|$ (including $B^+\to\tau^+\nu$), $\gamma$, and
$\gamma(\alpha)=\pi-\beta_{c\bar c}-\alpha$ are not affected by
the NP contribution and can be used in a (SM+NP) global fit to fix
the SM CKM parameters. We also consider only 3 generations of
quarks. The oscillation parameters, the weak phases, the
semi-leptonic asymmetries and the $B$-meson lifetime differences
are affected by the phase and/or the amplitude of the NP
contribution and allow to constrain the NP deviation to SM
quantified through out the model-independent parametrization:
$\langle B_{q}|M_{12}^{\textrm{\tiny SM+NP}}|\bar B_{q}\rangle =
\Delta_q^{\textrm{\tiny NP}} \langle B_{q}|M_{12}^{\textrm{\tiny
SM}}|\bar B_{q}\rangle$.

\vspace{-0.4cm}
\begin{figure}[h]
\begin{center}
{\hspace{-1cm}
\epsfig{file=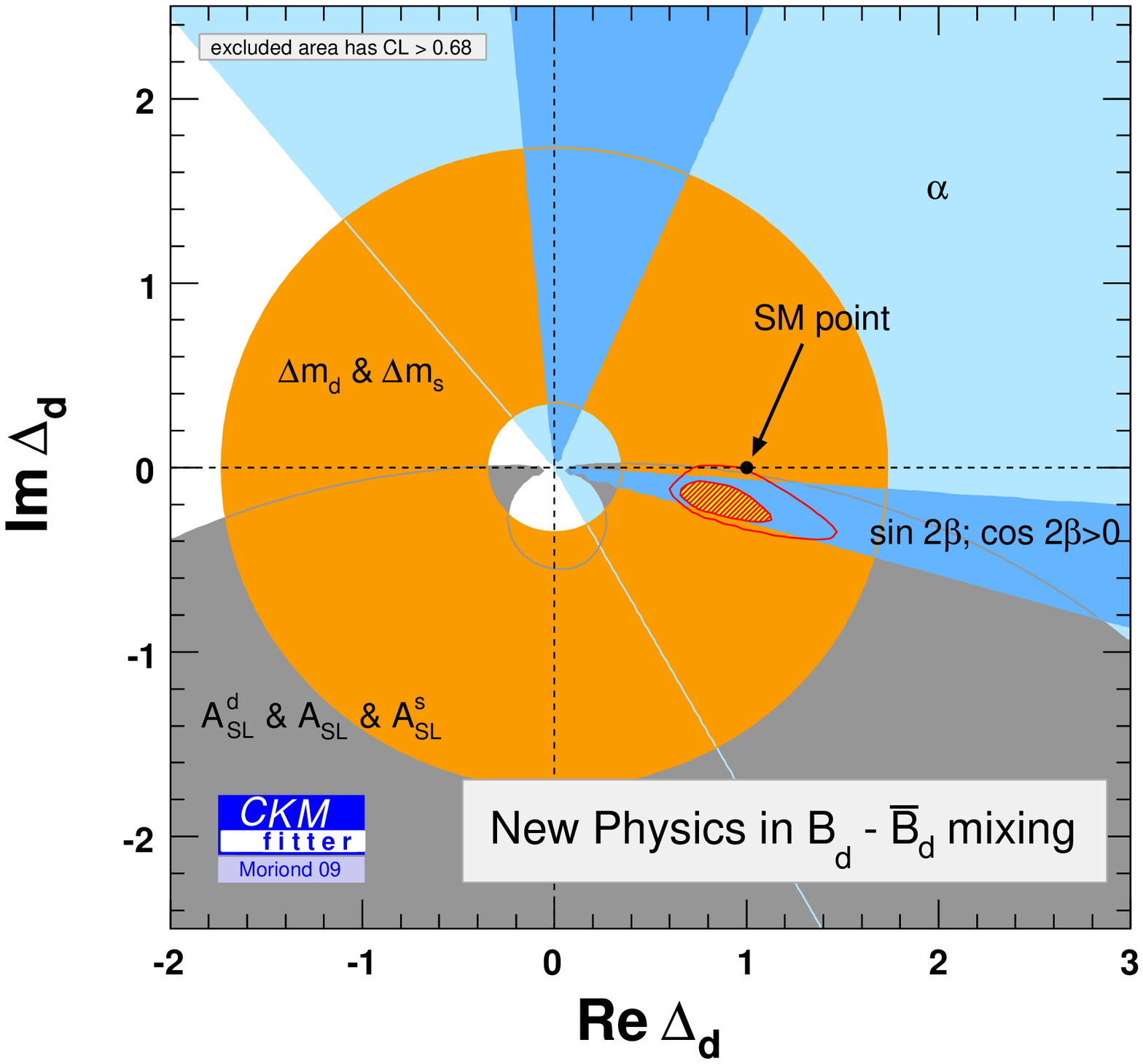,height=7.0cm,width=0.5\linewidth}}
{\hspace{-0.5cm}
\epsfig{file=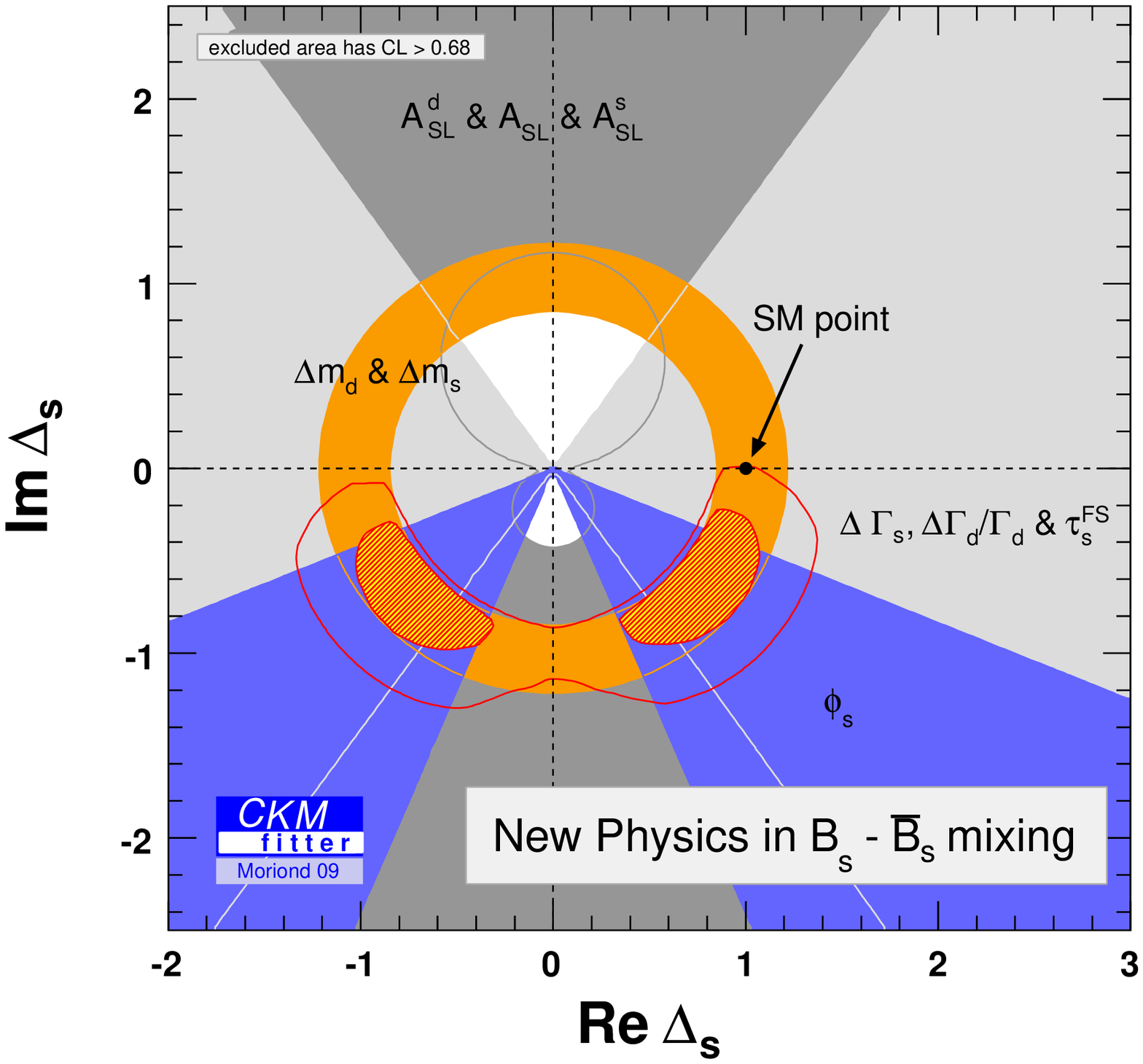,height=7.0cm,width=0.5\linewidth}}
\vspace{-0.25cm} \caption{68~\% CL contours for
$\Delta_q^{\textrm{\tiny NP}}$ in $B_d - \bar B_d$ system (Left)
and  in $B_s - \bar B_s$ system (Right). \label{fig:NP}}
\end{center}
\end{figure}

In Figure~\ref{fig:NP} we present the deviations to the SM
($\Delta_q^{\textrm{\tiny NP}}=1$) using the intuitive Cartesian
coordinates parametrization~\cite{ref:NiersteLenz}:
$\Delta_q^{\textrm{\tiny NP}}=\left( {\textrm {Re}} + i \ {\textrm
{Im}} \right) \Delta_q^{\textrm{\tiny NP}}$. This parametrization
is statistically more robust as uncertainties have Gaussian
behavior in the vicinity of $\vert \Delta_q^{\textrm{\tiny NP}}
\vert=0$. In the $B_d$ case, the tension in between $\sin(2\beta)$
and $|V_{ub}|_{\tau \nu}$ pushes the best fitted
$\Delta_d^{\textrm{\tiny NP}}$ 2.1~$\sigma$ away from the SM point
(while it is only 0.6~$\sigma$ away when $B^+\to\tau^+\nu$ is
removed). In the case of $B_s$, the deviation is 1.9~$\sigma$,
it's mainly driven by the recent TeVatron measurements of
$(2\beta_s,\Delta \Gamma_s)$~\cite{ref:HFAG}. This measurement is
performed with the time dependent analysis of the decay $B_s\to
J/\psi \phi$. It deviates by 2.2~$\sigma$ from the SM expected
value. In both cases $\Delta m_{q=d,s}$ constrain the modulus
$\vert \Delta_{q=d,s}^{\textrm{\tiny NP}} \vert$ to be in the
vicinity of 1 or below. this is the evidence of the KM mechanism
dominance for the sensitivity to NP effects. If one tests the
Minimal Flavor Violation (MFV) scenario (i.e. no additional NP
phase and Yukawa couplings only: ${\textrm {Im}}
\Delta_q^{\textrm{\tiny NP}}=0$ and $\Delta_d^{\textrm{\tiny
NP}}=\Delta_s^{\textrm{\tiny NP}}$), no tension with respect to SM
is observed, as theses effects arise at the present time only
through EW phases: $\sin(2\beta)$ vs. $|V_{ub}|_{\tau \nu}$ and
$\phi_s$, in both $B_{q=d,s}$ systems.

To conclude the KM mechanism is at work and dominates the
sensitivity to CP violation and to  NP in the b quark sector.
Anyway there is still substantial room for NP both in $B_d$-meson
and $B_s$-meson physics. Some few deviations to the SM global fit
exist at the present time and at most at the 2~$\sigma$ level. It
is therefore fundamental to finalize the analyzes of the present
B-factory datasets and to wait for the next generation experiments
at the LHC (huge $b$ quark cross-section production), or at the
future super-B factories, at KEK and possibly at Frascati (${\cal
L}=10^{35-36} cm^{-2}s^{-1}$). They will allow for high precision
measurements of rare effects. Finally, continuous progress in LQCD
are currently achieved, but even more accurate calculations, in a
coherent motion of that community, are mandatory and expected to
fully exploit the potential of the physics program in that field.

\section*{Acknowledgments}
I would like to thank all the members of the CKMfitter group for
the fruitful and stimulating collaboration in the preparation  on
the various topics covered here.

\end{document}